\documentclass[english,twoside,twocolumn]{IEEEtran}

\usepackage{amsfonts}
\usepackage{amsmath}   

\usepackage{amssymb}   
\usepackage{amscd}
\usepackage{babel}
\usepackage{latexsym}  
\usepackage{epsfig}
\usepackage{bbm}

\newtheorem{defi}{Definition}
\newtheorem{lemma}[defi]{Lemma}
\newtheorem{thm}[defi]{Theorem}

\newtheorem{exempel}[defi]{Example}

\DeclareRobustCommand\openone{\leavevmode\hbox{\small1\normalsize\kern-.33em1}}%

\newcommand{\qed}{\hfill $\blacksquare$}
\newcommand{\tr}{{\operatorname{Tr}}}
\newcommand{\id}{{\operatorname{id}}}

\newcommand{\bra}[1]{{\langle{#1}|}}
\newcommand{\ket}[1]{{|{#1}\rangle}}
\newcommand{\ketbra}[1]{{\ket{#1}\!\bra{#1}}}
\newcommand{\C}{{\mathbb{C}}}
\newcommand{\R}{{\mathbb{R}}}

\newcommand{\fset}[1]{{\mathcal{#1}}}

\newcommand{\1}{{\openone}}

\newlength{\blank}
\newenvironment{beweis}[1][{\hspace{-\blank}}]{{\noindent\emph{Proof\hspace{\blank}{#1}.\ }}}
{\hfill $\Box$\vskip 0.5\baselineskip}

\begin{document}

\title{Classical Capacity of\protect\\ Quantum Binary Adder Channels}
\author{Gleb V. Klimovitch$^*$\thanks{$^*$Department of Electrical Engineering, Stanford University, CA 94305, USA, and Watkins--Johnson Company, Stanford Research Park, 3333 Hillview Ave, Palo Alto, CA 94304--1223, USA. Email: {\tt gleb@stanford.edu}.}
and Andreas Winter$^\bullet$\thanks{$^\bullet$Was with Department of Computer Science, University of Bristol. Current address: Department of Mathematics, University of Bristol, Bristol BS8 1TW, United Kingdom. Email: {\tt a.j.winter@bris.ac.uk}.}
\thanks{An earlier version of this work was presented by the first author at the IEEE Symposium on Information Theory 2001, Washington DC, June 24--29, 2001, Proc.~ISIT 2001, p.~278.}
\thanks{Version of 9 February, 2005.}
}
\date{9 February 2005}

\maketitle

\begin{abstract}
  We analyze the quantum binary adder channel, i.e.~the quantum generalization of
  the classical, and well--studied, binary adder channel: in this model qubits
  rather than classical bits are transmitted. This of course is as special case
  of the general theory of quantum multiple access channels, and we may
  apply the established formulas for the capacity region to it.
  However, the binary adder channel is of particular interest classically,
  which motivates our generalizing it to the quantum domain. It turns out to
  be a very nice case study not only of multi--user quantum information theory,
  but also on the role entanglement plays there. It turns out that
  the analogous classical situation, the multi--user channel supported by
  shared randomness, is not distinct from the channel without shared randomness,
  as far as rates are concerned. However, we discuss the effect the new resource
  has on error probabilities, in an appendix.
  \par
  We focus specially on the effect entanglement between the senders as well
  as between senders and receiver has on the capacity region. Interestingly,
  in some of these cases one can devise rather simple codes meeting the capacity
  bounds, even in a zero--error model, which is in marked difference
  to code construction in the classical case.
\end{abstract}


\section{Classical and quantum\protect\\ binary adder channels}
\label{sec:adder}
The binary adder channel is a popular and well--studied example of
a multiple access channel in classical information theory:
$L$ senders may each choose a bit $x_i\in\{0,1\}$, which results in the
receiver getting
$$y=x_1+\ldots+x_L\in\{0,\ldots,L\}.$$
I.e.,~the receiver can have only very limited information on the
sent bits: e.g.~if $y=1$ she knows that exactly one $x_i$ equals
$1$, the others being $0$, but has no information on $i$.
It is easily seen that this channel (which remarkably is
deterministic) is equivalent to the channel randomly permuting
the bits $(x_1,\ldots,x_L)$: in one direction, from such a random permutation
the receiver can still calculate the sum $y$ of the bits, thus
simulating the output of the former channel. In the other direction,
$y$ can be used to generate a uniform distribution on the words
$(x_1,\ldots,x_L)$ with weight $y$, which is a simulation
of the latter channel.
\par
The general expression for the capacity region of multiple access
channels, as determined by Ahlswede~\cite{ahlswede:mac,ahlswede:mac2},
can be evaluated explicitly (see e.g.~\cite{cover:thomas}),
and gives for the two--user case $L=2$ (to which we shall
restrict our attention for the moment) the achievable
rate pairs $(R_1,R_2)$ of non--negative reals $R_1$, $R_2$ with
\begin{equation}
  \label{eq:classical:adder:cap}
  R_1,R_2\leq 1,\quad R_1+R_2\leq \frac{3}{2},
\end{equation}
for asymptotic block coding. This result is obtained by random
coding arguments and it is still an open problem to construct
codes achieving these bounds. Especially the zero--error case
received much attention, and we refer to the
survey~\cite{khachatrian:adder}, and to~\cite{ahlswede:balakirsky}
as the most recent contribution.
There is also a literature on low--error codes: see e.g.~\cite{hughes:cooper}.
\par
The above reasoning on the adder channel makes it plausible
to define the \emph{quantum binary adder channel} as follows:
it has inputs $L$ qubits, i.e.~states on two--dimensional Hilbert
spaces ${\cal H}_\ell\simeq\C^2$, $\ell=1,\ldots,L$, and acts as random permuter
of these qubits (for thoughts on the general methodology of quantum
information theory we refer the reader to~\cite{bennett:shor}).
Formally, define for a permutation $\pi\in S_L$ the
permuting operators on ${\cal H}=\bigotimes_{\ell=1}^L {\cal H}_\ell$ by
$$F_\pi:\ket{\psi_1}\otimes\cdots\otimes\ket{\psi_L}
          \longmapsto \ket{\psi_{\pi(1)}}\otimes\cdots\otimes\ket{\psi_{\pi(L)}},$$
and let the adder channel $\alpha$ be the following completely
positive, trace preserving (\emph{c.p.t.p.}) map on ${\cal B}({\cal H})$:
\begin{eqnarray*}
  \label{eq:q:adder:def}
  \alpha: & \sigma                               & \longmapsto
                                             \frac{1}{L!}\sum_{\pi\in S_L} F_\pi \sigma F_\pi^* \\
  \nonumber
          & \sigma_1\otimes\cdots\otimes\sigma_L & \longmapsto
                 \frac{1}{L!}\sum_{\pi\in S_L} \sigma_{\pi(1)}\otimes\cdots\otimes\sigma_{\pi(L)}.
\end{eqnarray*}
To send classical information via this channel the senders will choose input qubits
to their systems, while the receiver will choose a measurement, described by
a positive operator valued measure (POVM).
For example the senders could choose to send only states from
the fixed basis $\ket{0},\ket{1}\in\C^2$, and the receiver
performing the von Neumann measurement consisting
of the projectors
$$\ketbra{x_1\ldots x_L}=\ketbra{x_1}\otimes\cdots\otimes\ketbra{x_L}.$$
This obviously reproduces the behaviour of the classical adder channel,
making $\alpha$ a generalization of the former.
\par
However, we shall be concerned also with
the effect of entanglement on the transmission capacity of this channel:
in this case we assume that the senders and the receiver share initially
some multipartite entangled state, and sending information is
by the senders modifying their respective share of this state
$\ket{\iota}\in\bigotimes_{\ell=1}^L {\cal K}_\ell \otimes {\cal H}_R$
by applying quantum operations
(i.e., c.p.t.p.~maps) and subsequently putting it into $\alpha$.
\par
The details of these procedures are discussed more precisely below,
but we can remark that the channel proposed is an
example of a \emph{quantum multiple access channel}:
the first who appeared to have discussed the model are Allahverdyan and
Saakian~\cite{allah:saa:qmac}. The capacity region in full was determined
in~\cite{winter:qmac} (the result being reproduced in~\cite{hzh:qmac} for
the particular case of \emph{pure} signal states), in
the model of \emph{product state encodings} (i.e. the same condition under
which the Holevo bound holds and is achieved with single--user
channels~\cite{holevo:coding}).
The result, for the two--sender case to which we shall restrict ourselves
from here on, is as follows:
Suppose user $1$ may take actions $i\in\fset{I}$, user $2$ actions $j\in\fset{J}$,
which results in the (possibly mixed) output state $W_{ij}$ on Hilbert
space ${\cal H}$. This is a very
general description of a quantum multiple access channel, which obviously includes
the ones discussed above (with or without entanglement). We assume that
the channel acts \emph{memoryless}, meaning that in $n$ uses of the channel,
with inputs $i^n=i_1\ldots i_n\in\fset{I}^n$ and
$j^n=j_1\ldots j_n\in\fset{J}^n$ the output state will be
$$W^n_{i^nj^n}=W_{i_1j_1}\otimes\cdots\otimes W_{i_nj_n}
                                        \text{ on }{\cal H}^{\otimes n}.$$
\par
An \emph{$(n,\lambda)$--block code} for this channel is defined as a
triple $(f_1,f_2,D)$, with two functions
\begin{align*}
  f_1: &\ {\cal M}_1\longrightarrow \fset{I}^n, \\
  f_2: &\ {\cal M}_2\longrightarrow \fset{J}^n,
\end{align*}
($\fset{M}_1$, $\fset{M}_2$ being finite sets of messages), and a
decoding POVM $D=\left(D_{m_1m_2}\right)_{m_i\in\fset{M}_i}$, such that
the (average) error probability
\begin{equation*}\begin{split}
  e&(f_1,f_2,D) \\
   &\phantom{==}=1-\frac{1}{|\fset{M}_1| |\fset{M}_2|} \sum_{m_i\in\fset{M}_i}
                             \!\tr\!\left( W^n_{f_1(m_1)f_2(m_2)} D_{m_1m_2} \right)
\end{split}\end{equation*}
is at most $\lambda$.
The \emph{capacity region} ${\bf R}$ is then defined as the set of
all pairs $(R_1,R_2)$ such that there exist $(n,\lambda)$--block codes 
with the error probability $\lambda$ tending to zero, and the
code rates tending to $R_1$ and $R_2$, respectively, as $n\rightarrow\infty$:
$$\frac{1}{n}\log|\fset{M}_1| \longrightarrow R_1,
     \quad \frac{1}{n}\log|\fset{M}_2| \longrightarrow R_2.$$
Note that in this paper $\log$ is the logarithm to basis $2$.
\par
We will assume that ${\cal H}$ is finite,
so we might take $\fset{I}$, $\fset{J}$ to be finite, too. However, allowing general measure
spaces and measures $P$ on $\fset{I}$, $Q$ on $\fset{J}$, and a measurable map $W$ does
not change the result, but allows greater flexibility.
\begin{thm}
  \label{thm:qmac:cap}
  Denote by ${\bf R}_{PQ}$
  the set of points $(R_1,R_2)\in\R^2$ such that $R_1,R_2\geq 0$ and
  \begin{align*}
    R_1     &\leq I(P;W|Q),\\
    R_2     &\leq I(Q;W|P),\\
    R_1+R_2 &\leq I(P\times Q;W).
  \end{align*}
  Then the capacity region of the channel is given by the
  closed convex hull of the union of the ${\bf R}_{PQ}$.
  \qed
\end{thm}
\par
Here the information terms are quantum, as follows:
$$I(R;W)=H\left(\int dR(ij) W_{ij}\right)-\int dR(ij) H(W_{ij}),$$
with the von Neumann entropy $H$, and
$$I(P;W|Q)=\int dQ(j) I(P;W_{\cdot j}),$$
where $W_{\cdot j}$ is the single--user classical--quantum channel
conditional on $j$:
$$W_{\cdot j}:i\longmapsto W_{ij},$$
and likewise $W_{i\cdot}$ and $I(Q;W|P)$.
\par
Observe the formal analogy of this formula to the classical case, where
there appear mutual information and conditional mutual information,
too~\cite{ahlswede:mac,ahlswede:mac2}.
\par
We will use this formula to prove in the sequel (section~\ref{sec:noent})
that the capacity region of $\alpha$, with no entanglement available,
coincides with the region for the classical two--adder channel,
described by eq.~(\ref{eq:classical:adder:cap}). This we shall take
as the final piece of evidence that our definition really
represents \emph{the} quantum generalization of the classical adder channel.
Then we add entanglement to our investigation: in
section~\ref{sec:uuent} the enlargement of the capacity region due to
entanglement between the senders is investigated, while we allow
sender--receiver entanglement in section~\ref{sec:uurent},
increasing the capacity region once more, the latter effect
of course being reminiscent of dense coding~\cite{bennett:wiesner}.
To explain, however, the increase of the capacity due to entanglement
between the senders, we have to understand the particular kind of correlation
provided by it: in this direction, we discuss in the appendix
the easy fact that shared randomness between all of the parties
does not increase the capacity region of the classical adder channel
(in fact, this is even true for the quantum adder channel $\alpha$). So,
the observed increase of the capacity has to be attributed
to quantum effects.
\par\medskip
To end this introduction, a few words on previous and
related work: in~\cite{hzh:qmac}, final section, some remarks regarding
entanglement between the users are made. However, as
this paper is only concerned with the pure state case of multiple access
coding, there is no overlap with the present work.
\par
Two further works have come to our attention that touch upon
the peculiar ``interference'' (mutual disturbance)
between messages in a multiple access channel, both in a
situation where previous entanglement between two senders
and the receiver is assumed, and a noiseless channel is
considered (instead of our noisy random permuter):
In~\cite{gzty}, rather unaware of the information theoretic
meaning, the case of 1--ebit of sender--receiver entanglement in the
form of a GHZ--state is treated, in a noiseless setting:
in section 3.2. of that work it is shown that the rate--sum
$3$ is optimal. There is overlap with this work concerning the idea
of generalized superdense coding, compare subsection~\ref{subsec:1ebit}.
\par
In~\cite{liu:long:tong:li} this investigation is carried to $d$--level
$N$--party higher GHZ--states, and coding methods meeting the
capacity region bounds (which can be derived from
theorem~\ref{thm:qmac:cap}) are discussed.

\section{Two--user quantum adder channel}
These are the channels we are going to investigate in the sequel:
\\
Fix an initial pure state $\ket{\iota}$ of the system
${\cal H}_1\otimes{\cal H}_2\otimes{\cal H}_R$, where
${\cal H}_1={\cal H}_2=\C^2$ are the two users' qubit systems
(with fixed orthonormal basis $\ket{0}$, $\ket{1}$),
and ${\cal H}_R$ is the receiver's system (one may obviously assume that
${\cal H}_R=\C^4$, as the initial state is always pure).
As sets of allowed actions we define all local quantum operations:
$$\fset{I}=\fset{J}
       =\{\varphi:{\cal B}(\C^2)\rightarrow{\cal B}(\C^2)|\varphi\text{ c.p.t.p.}\}.$$
The channel $\alpha$ for two senders has the simple form
$$\alpha:\sigma\longmapsto \frac{1}{2}\left( \sigma+F\sigma F^*\right),$$
with the flip operator
$$F=F_{(12)}:\ket{u}\otimes\ket{v} \longmapsto \ket{v}\otimes\ket{u}.$$
Notice that a most convenient eigenbasis of this unitary
is provided by the \emph{Bell states}
\begin{align*}
  \ket{\Phi^+} &=\frac{1}{\sqrt{2}}(\ket{00}+\ket{11}), \\
  \ket{\Phi^-} &=\frac{1}{\sqrt{2}}(\ket{00}-\ket{11}), \\
  \ket{\Psi^+} &=\frac{1}{\sqrt{2}}(\ket{01}+\ket{10}), \\
               &                                        \\
  \ket{\Psi^-} &=\frac{1}{\sqrt{2}}(\ket{00}+\ket{11}),
\end{align*}
the first three (which span the \emph{symmetric subspace} ${\cal S}$)
with phase $+1$, the last with phase $-1$.
From this one can see that the effect of $\alpha$ is to destroy coherence
between ${\cal S}$ and $\C\ket{\Psi^-}$: it is equivalent to
an incomplete nondemolition von Neumann measurement of the
projector onto ${\cal S}$ and its complement $\ketbra{\Psi^-}$.
\par
The swap super--operator for density operators is defined
as follows: for a density operator $\sigma$ on $\C^2\otimes\C^2$ let
$$S(\sigma)=F\sigma F^*.$$
This means that the channel ``quantum binary adder with prior entanglement
$\ket{\iota}$'' is described by mapping
$f\in\fset{I}$, $g\in\fset{J}$ to the output state
$$W_{fg}=\frac{1}{2}\bigl(f\otimes g\otimes\id(\ketbra{\iota})+
                             (S\otimes\id)(f\otimes g\otimes\id(\ketbra{\iota}))\bigr).$$
\par\medskip
Making indistinguishable permutations of the input qubits surely is a necessary
requirement for a candidate quantum adder channel, as well as reducing to
the classical binary adder for the particular choice of input bases and
output measurement (both of which $\alpha$ satisfies).
Notice however that $\alpha$ even keeps coherence in the symmetric subspace
${\cal S}$.
Just as well one could destroy it by doing a nondemolition measurement
in this or some other basis after $\alpha$. Nevertheless, apart from being
hard to motivate (which basis to choose?), this is an unnecessary
``classicalisation'' of the channel: as we shall see in the next section 
our definition of quantum adder channel has the same capacity region
as the classical adder channel. We might take this as saying that $\alpha$
is the ``most quantumly'' channel generalizing the usual binary adder channel
and at the same time not increasing the capacity region.

\section{No entanglement}
\label{sec:noent}
Here we treat the case of a trivial receiver's system ${\cal H}_R=\C$,
and $\ket{\iota}=\ket{00}$. Thus coding amounts to independent choices
of states $\ket{\phi}$ and $\ket{\psi}$, and $\fset{I}$, $\fset{J}$ may be identified with
the sets of pure states on ${\cal H}_1,{\cal H}_2$, respectively (note that
choosing mixed states to encode is obviously suboptimal here).
We will show that the capacity region in this case is identical to the
classical adder channel's. To do this, we obviously have only to prove
that our quantum channel obeys the same upper rate bounds as the classical one
(because the classical coding schemes work identically for the quantum channel).
\par
Because the individual bounds on $R_1$ and $R_2$ are convex combinations of
quantities trivially upper bounded by $1$, we have only to show that
$R_1+R_2\leq 3/2$, which in turn will follow if we show that
$$I(P\times Q;W)\leq \frac{3}{2},$$
for all distributions $P$ and $Q$ on the pure qubit states.
In more extensive writing this means
\begin{equation}
  \label{eq:1}
  \begin{split}
  H&\!\left(\int\!\! dP(\phi)dQ(\psi)\frac{1}{2}(\ketbra{\phi}\!\otimes\!\ketbra{\psi}
                             +\ketbra{\psi}\!\otimes\!\ketbra{\phi})\right)              \\
   &-\int\!\! dP(\phi)dQ(\psi) H\!\left(\!\frac{1}{2}(\ketbra{\phi}\!\otimes\!\ketbra{\psi}
                             +\ketbra{\psi}\!\otimes\!\ketbra{\phi})\!\right)            \\
   &\phantom{============================}\leq \frac{3}{2}.
  \end{split}
\end{equation}
It is an easy exercise to show that for two vectors $\ket{v},\ket{w}$
with $t=|\bra{v}w\rangle|$, one has
\begin{equation}
  \label{eq:2}
  H\left(\frac{1}{2}(\ketbra{v}+\ketbra{w})\right)
    =H\left(\frac{1-t}{2},\frac{1+t}{2}\right),
\end{equation}
with the Shannon entropy of a binary distribution at the right hand side.
\par
Applied to the terms in the second integral in eq.~(\ref{eq:1}) we get
$$H\!\left(\! \frac{1}{2}(\ketbra{\phi}\!\otimes\!\ketbra{\psi}
                          +\ketbra{\psi}\!\otimes\!\ketbra{\phi}) \!\right)
  =H\!\left(\! \frac{1\pm |\bra{\phi}\psi\rangle|^2}{2} \!\right)\!.$$
Introducing
$$\rho_P=\int dP(\phi) \ketbra{\phi},\quad
  \rho_Q=\int dQ(\psi) \ketbra{\psi}$$
we can rewrite the left hand side of eq.~(\ref{eq:1}) as
\begin{equation}\begin{split}
  \label{eq:3}
  H&\left(\frac{1}{2}\rho_P\otimes\rho_Q+\frac{1}{2}\rho_Q\otimes\rho_P\right)        \\
   &\phantom{======}
            -\int dP(\phi)dQ(\psi)H\left(\frac{1\pm |\bra{\phi}\psi\rangle|^2}{2}\right).
\end{split}\end{equation}
Now an important observation comes in: the function
$H\left(\frac{1\pm x}{2}\right)$, for $0\leq x\leq 1$,
is strictly concave and strictly decreasing, with value $1$ at
$x=0$ and value $0$ at $x=1$. In particular
$$H\left(\frac{1-x}{2},\frac{1+x}{2}\right)\geq 1-x.$$
But Taylor expansion shows even more:
\begin{equation}
  \label{eq:4}
  1-x^2\leq H\left(\frac{1-x}{2},\frac{1+x}{2}\right)\leq 1-\frac{1}{2}x^2.
\end{equation}
Plugging this in we can lower bound the subtraction term in
eq.~(\ref{eq:3}) by
$$1-\int dP(\phi)dQ(\psi) |\bra{\phi}\psi\rangle|^2
   =1-\tr(\rho_P\rho_Q).$$
Thus we get an upper bound on the left hand side of eq.~(\ref{eq:1}):
\begin{equation}
  \label{eq:upper:noent}
  H\left(\frac{1}{2}\rho_P\otimes\rho_Q+\frac{1}{2}\rho_Q\otimes\rho_P\right)
                                                            -1+\tr(\rho_P\rho_Q).
\end{equation}
The maximum of this expression is obtained when $\rho_P$ and $\rho_Q$ commute,
in fact if they are equal: replacing both $\rho_P$ and $\rho_Q$ by
$\frac{1}{2}(\rho_P+\rho_Q)$ increases both the entropy contribution
(because of subadditivity), and the trace contribution:
$$\tr\left(\frac{\rho_P+\rho_Q}{2}\right)^2-\tr\rho_P\rho_Q
    =\tr\left(\frac{\rho_P-\rho_Q}{2}\right)^2\geq 0.$$
But with commuting $\rho_P$, $\rho_Q$ the situation is essentially the
classical one, and we are done. More precisely, $P=Q$ may be taken
as distribution on a common eigenbasis $\ket{0},\ket{1}$ of
$\rho_P=\rho_Q$, in which case the maximum in eq.~(\ref{eq:upper:noent})
is easily seen to be attained at $\rho_P=\frac{1}{2}\1$: when
$\rho_P=\rho_Q$ the expression in eq.~(\ref{eq:upper:noent}) becomes
$$2H(\rho)-1+\tr\left(\rho^2\right).$$
In terms of $\rho$'s eigenvalues $(1\pm y)/2$ this reads, and
can be estimated, as
$$2H\left(\frac{1-y}{2},\frac{1+y}{2}\right)+\frac{y^2-1}{2}
    \leq \frac{3}{2}-\frac{1}{2}y^2,$$
the latter clearly obtaining the maximum at $y=0$.
\par
Observe that in this case only $\langle\phi\ket{\psi}=0$
or $=1$ occur with positive probability in eq.~(\ref{eq:3}), so in
eq.~(\ref{eq:upper:noent}) we have in fact \emph{equality}: the points
in this region can be achieved by using the classical input states
$\ket{0},\ket{1}$, the corresponding POVM $\bigl(\ketbra{xy}:x,y\in\{0,1\}\bigr)$
followed by a classical postprocessing (decoding). Observe that
we cannot, however, provide explicit code constructions
to do this, as was pointed out in the introduction.

\section{Sender--sender entanglement}
\label{sec:uuent}
Up to basis change the most general two--qubit (pure) state that can be
shared among the senders is
$$\ket{\iota}=\alpha\ket{00}+\beta\ket{11},$$
with $\alpha\geq\beta\geq 0$ and $\alpha^2+\beta^2=1$.
\par
Before we go into the general case, we consider the two extremes:
\par\noindent
\emph{1. No entanglement}:
$\alpha=1$ (meaning no entanglement) was treated in the previous
section, and the capacity region determined.
\par\noindent
\emph{2. Maximal entanglement}:
on the other hand, for $\alpha=\beta=1/\sqrt{2}$ (maximal entanglement),
the upper bounds from theorem~\ref{thm:qmac:cap} are trivially
bounded by $2$, each; $R_1,R_2,R_1+R_2\leq 2$. As it turns out,
this \emph{is} the capacity region. For example, the corner
$(2,0)$ is achieved by sender two sending nothing, while sender
one modulates $\ket{\iota}$ as in dense coding. Note that the
four Bell--states are invariant under the channel, so the
$2$ bits encoded by sender one are recovered without error.
This is a notable feature, as we pointed out the difficulty
of finding error--free codes for the unassisted adder channel.
\par
Now we approach the general case:
one strategy which seems to be good is to (asymptotically
reversibly!) concentrate
the $n$ copies of the state $\ket{\iota}$ into
$k = n H(\alpha^2,\beta^2) - o(n)$ many EPR pairs~\cite{concentrate}.
Then use time sharing between $k$ uses of the
maximal entanglement scheme (item 2 above) and
$n-k$ uses of the no entanglement scheme (item 1 above),
resulting in an achievable rate region cut out by
the inequalities
\begin{align}
  \label{eq:uue-R}
  R_1, R_2 &\leq 2 \cdot H(\alpha^2,\beta^2)
               + 1 \cdot \bigl( 1-H(\alpha^2,\beta^2) \bigr) \nonumber \\
           &=    1+H(\alpha^2,\beta^2), \\
  \label{eq:uue-RR}
  R_1+R_2  &\leq 2 \cdot H(\alpha^2,\beta^2)
               + \frac{3}{2} \cdot \bigl( 1-H(\alpha^2,\beta^2) \bigr) \nonumber \\
           &=    \frac{3}{2}+\frac{1}{2}H(\alpha^2,\beta^2).
\end{align}
The right hand side of eq.~(\ref{eq:uue-R}), $1+H(\alpha^2,\beta^2)$
is easily seen to be actually an upper bound on any achievable
individual rate: indeed, let the second sender cooperate optimally,
by sending all his entanglement to the receiver. Then, even
disregarding the channel noise, the maximal entanglement
between the first sender and the receiver is $n H(\alpha^2,\beta^2)$,
and it is fairly easy to show that under these circumstances
sending $n$ qubits (again disregarding the noise)
can transmit at most an asymptotic rate of $1+H(\alpha^2,\beta^2)$
classical bits~\cite{hiroshima,winter:spqg+add}.
\par
In view of this, we also conjecture that
the right hand side of eq.~(\ref{eq:uue-RR}), $\frac{3}{2}+\frac{1}{2}H(\alpha^2,\beta^2)$,
is always an upper bound on the rate sum. A proof of this, however,
has eluded us so far.

\section{Sender--receiver entanglement}
\label{sec:uurent}
We shall study two cases of entanglement between senders and receiver,
both distinguished by their symmetry: the case of a shared GHZ--state
in subsection~\ref{subsec:1ebit}, and the case of maximal entanglement
in subsection~\ref{subsec:2ebits}.

\subsection{1 ebit}
\label{subsec:1ebit}
Here the parties share initially a GHZ--state
$$\ket{\iota}=\frac{1}{\sqrt{2}}(\ket{000}+\ket{111}).$$
\par
Note that this is the unique three--qubit state (up to local unitaries)
having all its single particle states equal to the maximally mixed state.
\par
Now, we shall prove that the region described by the inequalities
$$R_1,R_2\leq 2,\quad R_1+R_2\leq \frac{5}{2},$$
is indeed the full capacity region: the individual rate bounds are obvious
again, so we have only to bound the rate--sum.
\par
We begin again with considering the unitary case:
Let the users employ unitaries
\begin{equation*}
  U_a=\left(\begin{array}{rr}
              \alpha & -\overline{\beta} \\
              \beta  &  \overline{\alpha}
            \end{array}\right),\quad
  U_b=\left(\begin{array}{rr}
              \gamma & -\overline{\delta} \\
              \delta &  \overline{\gamma}
            \end{array}\right).
\end{equation*}
(Global phases do not matter).
With the flip unitary $F$ from above, it is a straightforward calculation
to obtain
\begin{equation*}
  \begin{split}
  \left|\bra{\iota}U_a^*\otimes U_b^*\otimes\1
            (F\otimes\1)
            U_a\otimes U_b\otimes\1\ket{\iota}\right|
                               &=|\bra{0}U_a^* U_b\ket{0}| \\
                               &=|\bra{1}U_a^* U_b\ket{1}|.
  \end{split}
\end{equation*}
Thus we can estimate (with $T_{ab}=U_a\otimes U_b\otimes\1$)
\begin{equation*}\begin{split}
  &R_1+R_2                                                                    \\
  &\phantom{/}
   \leq \! H\!\!\left(\int\!\! dP(a)dQ(b)\frac{1}{2}
                      \bigl(T_{ab}\ketbra{\iota}T_{ab}^*
                        \!+\! S\otimes\id(T_{ab}\ketbra{\iota}T_{ab}^*)\!\bigr)
                      \!\!\right)                                             \\
  &\phantom{==========}
   -\int\! dP(a)dQ(b)H\!\left(
                                   \frac{1\pm |\bra{0}U_a^* U_b\ket{0}|}{2}
                                           \right)                            \\
  &\phantom{/}
   \leq \! H\!\!\left(\int\!\! dP(a)dQ(b)\frac{1}{2}
                      \bigl(T_{ab}\ketbra{\iota}T_{ab}^*
                        \!+\! S\otimes\id(T_{ab}\ketbra{\iota}T_{ab}^*)\!\bigr)
                      \!\!\right)                                             \\
  &\phantom{==========}
   -1+\int\! dP(a)dQ(b) |\bra{0}U_a^* U_b\ket{0}|^2,
\end{split}\end{equation*}
where we have used eq.~(\ref{eq:4}).
We will employ the following important inequality:
\begin{lemma}
  \label{lemma:inequality}
  For a general state $\rho$ on the composite system
  ${\cal H}_1\otimes{\cal H}_2$ and a POVM $(E_1,\ldots,E_n)$
  on ${\cal H}_2$, it holds for the measurement probabilities
  $\lambda_i=\tr\bigl(\rho(\1\otimes E_i)\bigr)$ and the post--measurement states
  $$\sigma_i=\frac{1}{\lambda_i}
             \left(\sqrt{\1\otimes E_i}\right)\rho\left(\sqrt{\1\otimes E_i}\right)$$
  that
  $$H(\rho)\leq H(\lambda_1,\ldots,\lambda_n)
                   +\sum_{i=1}^n \lambda_i H(\sigma_i).$$
\end{lemma}
\begin{beweis}
  Defining
  $$\rho_i=\frac{1}{\lambda_i}\sqrt{\rho}\bigl(\1\otimes E_i\bigr)\sqrt{\rho},$$
  one has $\rho=\sum_j \lambda_j\rho_j$, and
  $H(\rho_j)=H(\sigma_j)$ for all $j$: in fact $\rho_j$ and $\sigma_j$ are
  conjugate operators via a unitary, by the polar decomposition. Finally, by the
  data processing inequality~\cite{ahlswede:loeber}
  $$H(\lambda)\geq H(\rho)-\sum_j \lambda_j H(\rho_j),$$
  and we are done.
\end{beweis}
\par
We apply this to the complete measurement in the basis $\{\ket{0},\ket{1}\}$
on the receiver's system ${\cal H}_R$, and obtain:
\begin{equation*}\begin{split}
  &R_1+R_2 \leq   \\
  &\phantom{=}
   1\!+\!\sum_{k=0}^1 \frac{1}{2}\left[
              H\!\Bigg(\!\int\! dP(a)dQ(b)
                       \frac{1}{2}\Big(U_a\!\otimes\! U_b\ketbra{kk}U_a^*\!\otimes\! U_b^*
                                                                      \right. \\
  &\phantom{==============:}
           \left. +S(U_a\!\otimes\! U_b\ketbra{kk}U_a^*\!\otimes\! U_b^*)\Big)
                                                                      \Bigg)\right. \\
  &\phantom{============:}
           -1+\int dP(a)dQ(b)|\bra{k}U_a^*U_b\ket{k}|^2\Bigg]\!.
\end{split}\end{equation*}
Each of the two terms corresponding to $k=0,1$ can be written in the form
\begin{equation*}\begin{split}
  H&\left(\int d\tilde{P}(\phi)d\tilde{Q}(\psi)
         \frac{1}{2}(\ketbra{\phi}\otimes\ketbra{\psi}
                     +\ketbra{\psi}\otimes\ketbra{\phi})
    \right)       \\
   &\phantom{==============}
    -1+\int d\tilde{P}(\phi)d\tilde{Q}(\psi) |\langle\phi\ket{\psi}|^2.
\end{split}\end{equation*}
Since this --- by the reasoning of section~\ref{sec:noent} --- is
bounded by $3/2$, we get the desired bound $R_1+R_2\leq 5/2$.
\par
Now we demonstrate how, utilizing a code for the classical adder channel,
the points in the region
$$R_1,R_2\leq 2,\quad R_1+R_2\leq\frac{5}{2}$$
can be achieved. For this, let an $n$--block code for the classical
binary adder channel be given, with rates $R_1$ and $R_2$.
This code can be used to encode using the initial GHZ--state:
instead of sending a bit $b$ each sender applies $\sigma_x^b$.
It is easily seen that by a C--NOT of his share of $\ket{\iota}$
onto the other two qubits he receives through the channel and
measuring in the standard basis the receiver obtains exactly the
same data as in the classical case with the classical adder code.
\par
However, this still allows sender one (say) to encode an extra bit
into the phase: she either applies $\sigma_z$ or $1$ on her
part of $\ket{\iota}$, thus either leaving the joint state
in the subspace
$${\cal P}={\rm span}\left\{\frac{1}{\sqrt{2}}(\ket{ab0}+\ket{\overline{a}\overline{b}1}):
                                                                   a,b\in\{0,1\}\right\},$$
or steering it isometrically into the orthogonal subspace
$${\cal N}={\rm span}\left\{\frac{1}{\sqrt{2}}(\ket{ab0}-\ket{\overline{a}\overline{b}1}):
                                                                   a,b\in\{0,1\}\right\}.$$
Note that both these subspaces are stable under the channel
and that the named $\sigma_z$ action commutes with it.
So, the receiver can decode the extra bit first, by distinguishing
first ${\cal P}$ and ${\cal N}$ by a nondemolition measurement
(then rotating back into ${\cal P}$ if need be) and then proceeding
as just described with the classical code.
This achieves the rate point $(1+R_1,R_2)$. Equally, we can
achieve the point $(R_1,1+R_2)$, proving our claim.

\subsection{2 ebits --- maximal entanglement}
\label{subsec:2ebits}
The general form of an entangled state between the
users' systems and the receiver's is
\begin{equation*}\begin{split}
  \ket{\iota} &=\alpha_0\ket{\Phi^+}\otimes\ket{u_0}+\alpha_1\ket{\Phi^-}\otimes\ket{u_1} \\
              &\phantom{=}
               +\alpha_2\ket{\Psi^+}\otimes\ket{u_2}+\alpha_3\ket{\Psi^-}\otimes\ket{u_3}.
\end{split}\end{equation*}
This to have maximal entanglement it is necessary and sufficient that
$\alpha_i=1/2$ for all $i$, and the $u_i$ are orthonormal vectors.
\par
In this case the general upper bounds for $R_1,R_2$ are dominated
by $R_1,R_2\leq 2$, so again we only have to estimate the rate--sum
$R_1+R_2$: first, for \emph{unitary} actions $U_1$, $U_2$ of both users,
the resulting state $U_1\otimes U_2\ketbra{\iota}U_1^*\otimes U_2^*$
is still maximally entangled, hence the output state is an equal
mixture of
\begin{equation*}
  \frac{1}{2}(\ket{\Phi^+}\otimes\ket{v_0}+\ket{\Phi^-}\otimes\ket{v_1}
              +\ket{\Psi^+}\otimes\ket{v_2}\pm\ket{\Psi^-}\otimes\ket{v_3}),
\end{equation*}
for an orthonormal system $v_0,\ldots,v_3$.
Its von Neumann entropy is, by equation~(\ref{eq:2}), equal to
$H\left(\frac{1}{4},\frac{3}{4}\right)$. Thus, if both users are restricted
to unitary codings, we find
$$R_1+R_2\leq 4-H\left(\frac{1}{4},\frac{3}{4}\right) \approx 3.189,$$
and the bound can be achieved by taking the normalized Haar measure for
either user, or --- more discretely --- the uniform distribution on
the Pauli unitaries (including $\1$) for either user.
Note that again, like in the case of the classical adder channel, we cannot give
an explicit good code, nor is it obvious to produce optimal zero--error codes:
our argument relies on the random coding in the general coding
theorem~\ref{thm:qmac:cap}.

\subsection{Significance of our findings}
To begin with, comparing the rate regions described so far,
we see that they increase as we go from no entanglement,
via partial to maximal sender--sender entanglement,
and further as we increase the sender--receiver
entanglement.
There are, however, two important caveats to consider
before concluding that the \emph{capacity region}
increases with the available entanglement
(apart from only conjecturing eq.~(\ref{eq:uue-RR})
to be the optimal bound on the rate sum for partial
sender--sender entanglement):
\par
First, we have in both scenarios considered
in section~\ref{sec:uurent} assumed that the
encoding is done using unitaries. Though we don't
expect non--unitary operations to perform better
(as they introduce further noise) we lack a proof of
that statement.
\par
Second, we have on occasions (sections~\ref{sec:noent}
and~\ref{sec:uurent}) only considered single--copy
maximisation of the involved informations.
As was pointed out in the beginning the outer bounds on the capacity region
apply in general only if the signal states allowed in coding are products.
This is satisfied in all situations under investigation here if the encoding
operations (unitary or general) are products themselves.
\par
The situation may change if arbitrary encodings on blocks are allowed
(as maybe for the single--user channel: this relates to the additivity
of the Holevo capacity of a channel (see~\cite{s:w}). It is quite
conceivable that the analysis of section~\ref{sec:noent} will not apply
any longer. The same holds for the entanglement--assisted situations: only
in a case like maximal entanglement in
section~\ref{sec:uuent} the bounds are so simple
that they obviously still apply here.
\par
We would like to point out that the problem of determining the capacity
region in the presence of entanglement (in particular to find out if it increases at
all beyond the separable encoding region discussed above) is not
contained in the analogous discussion of~\cite{winter:qmac}:
this is another and more general instance of the extended additivity
question for the classical capacity of quantum channels,
discussed in~\cite{winter:spqg+add}: notice that in the discussion there,
as in the case of section~\ref{sec:noent} above, the sender input
states of their choosing into the channel, while in the presence of
entanglement they input quantum operations.

\section{The case of many senders}
\label{sec:many:users}
Let us now turn to an investigation of the analogous problem
for $L>2$ senders, each sharing an ebit with the receiver initially:
\begin{equation*}
  \ket{\Phi} = 2^{-L/2} \sum_{x_1,\ldots,x_L}
                          \ket{x_1}\otimes\cdots\otimes\ket{x_L}
                                      \otimes\ket{x_1\ldots x_L}_R.
\end{equation*}
Instead of aiming at finding the whole capacity region of
the $L$--user quantum binary adder channel, we will
concentrate on a quantity of particular interest for
symmetric channels like this one: the maximal rate sum
$\Sigma R = R_1+\ldots+R_L$.
We remind the reader that for a classical binary adder
channel this is maximized by $L$ uniform input distributions,
for which $\Sigma R$ becomes the entropy of a binomial distribution,
and $\Sigma R \sim \frac{1}{2}\log L$ for large $L$~\cite{chang:weldon}.
\par
For the entanglement--assisted quantum case, consider for
the moment a scheme where each user modulates her share
of the entangled state $\ket{\Phi}$
with a Pauli operator, uniformly chosen at random
from $\{ \1,\sigma_x,\sigma_y,\sigma_z \}$. Because these
can always be commuted through to acting on $R$, all signal
states of this channel are (up to local unitaries on the receiver system)
equivalent to
$$\tau := \frac{1}{L!}\sum_\pi (F_\pi\otimes\1)\ketbra{\Phi}(F_\pi^*\otimes\1),$$
and the average output is maximally mixed on $2L$ qubits.
To evaluate the Holevo information in the bound for the rate sum,
we thus have to calculate the entropy of the state $\tau$.
\par
\begin{figure}[ht]
  \centering
  \includegraphics[width=8.5cm]{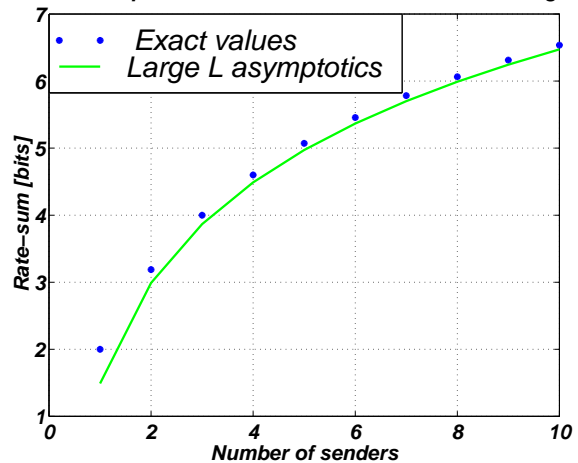}
  \caption{Achievable rate sum for the quantum binary adder channel with
    maximal prior entanglement between senders and receiver. The dots
    are the exact values from eq.~(\ref{eq:q-rate-sum}), while the line is the
    asymptotics of these values for large $L$,
    $\Sigma R \sim \frac{3}{2}\log L$.}
  \label{fig:manysenders}
\end{figure}
\par
Since $\tau$ is an average over the symmetric group, we write it in terms
of the Clebsch--Gordan decomposition of $(\C^2)^{\otimes L}$ into irreducible
decompositions of $U(2)$ and $S_L$~\cite{wigner}:
$$(\C^2)^{\otimes L}
    = \bigoplus_{k=0}^{\lfloor L/2 \rfloor} {\cal S}_k \otimes {\cal P}_k,$$
with the ($U(2)$, irreducible) spin representations ${\cal S}_k$ of dimension
$L-2k+1$, and the ($S_L$, irreducible) permutation representations
${\cal P}_k$ of dimension $d_k = {L \choose k} \!-\! {L \choose {k-1}}$.
\par
Hence, by writing $\ket{\Phi}$ in this decomposition for
both sender and receiver system, and applying Schur's lemma,
$$\tau = \bigoplus_{k=0}^{\lfloor L/2 \rfloor}
             p_k \ketbra{\Phi_k}_{{\cal S}_k{\cal S}_k}
                 \otimes \frac{1}{d_k}\1_{{\cal P}_k}
                 \otimes \frac{1}{d_k}\1_{{\cal P}_k},$$
with some maximally entangled state $\ket{\Phi_k}$ on
${\cal S}_k \otimes {\cal S}_k$.
From counting dimensions,
we can calculate the (probability!) weights in this sum:
$p_k = d_k (L-2k+1) 2^{-L}$, and we get,
\begin{equation}
  \label{eq:q-rate-sum}
  \Sigma R = 2L - H\bigl( p_0,\ldots,p_{\lfloor L/2 \rfloor} \bigr)
                - \sum_{k=0}^{\lfloor L/2 \rfloor} p_k \log\bigl( d_{k}^2 \bigr).
\end{equation}
For $L=2$ we recover our result from section~\ref{subsec:2ebits},
which gives rate sum $\approx\! 3.189$; for
$L=3$ and $4$, this formula gives values
$4$ and $\approx\! 5.057$, respectively.
For large $L$, we can quite straightforwardly
see that $\Sigma R \sim \frac{3}{2}\log L$.
The plot in figure~\ref{fig:manysenders} illustrates the result.

\section{Concluding remarks}
\label{sec:conclude}
We have introduced and studied the quantum binary adder channels,
determining its capacity region in the case of two senders,
without prior entanglement and with the help of various three--party
entangled states between the senders and the receiver. It turned out
that sender--sender entanglement already increases the capacity region
(and that this region is indeed directly related to the amount
of entanglement available),
to become even larger for sender--receiver entanglement,
which we studied in two important cases: a GHZ--state and maximal
entanglement (2 ebits).
\par
For a large number $L$ of users, we found that maximal
sender--receiver entanglement almost triples the achievable
rate sum compared the classical adder channel. Though we didn't
prove it, it seems likely that our figure actually is also
best possible.
\par
Among questions that deserve further study we would like to advertise
two as specially interesting: First, as the case of ``much entanglement'' in our
case study proved extremely fruitful, we are motivated to ask about
the \emph{entanglement assisted} capacity region of a quantum multiple access
channel, in the spirit of the beautiful work~\cite{bsst}, where the classical
capacity of a quantum channel was studied in the presence of
\emph{arbitrary} entanglement.
Second, we propose the problem of finding the rates of
\emph{quantum} information transmission
via the quantum adder channel, and more generally for an arbitrary quantum
multiple access channel, which lies out of the scope of the present
investigation.

\section*{Acknowledgments}
The authors would like to thank A.~S.~Holevo for his help in bringing
about their collaboration.

GVK was supported by the Watkins--Johnson Company.
AW was partially supported by the SFB 343 ``Diskrete Strukturen in der
Mathematik'' of the Deutsche Forschungsgemeinschaft, by the University
of Bristol, and by the U.K. Engineering and Physical Sciences Research
Council.

\section*{Appendix: Shared randomness in\protect\\ multiuser information theory}
\label{app:shared:random}
The classical analogue of entanglement between the communication parties
is shared randomness. Does this additional resource change capacity
regions?
\par
As it turns out, the answer is ``no'': the reason being that the use
of shared randomness can be described as (jointly) randomly using
several ordinary communication protocols. 
Also, in multiuser situations we favour the average error concept
(average error probabilities over assumed uniform distribution on
all message sets) over the familiar maximal error concept in
single--user situations. Hence, if we are given a code with
shared randomness and all its $K$ (average) error probabilities bounded by
$\epsilon$, there is one of the constituent ordinary
codes with average error probabilities bounded by $K\epsilon$.
Observe that $K$ is a constant of the setup, e.g.~the number
of senders in the multiple--access channel.
\par
So, allowing the use of shared randomness does not increase the capacity
regions.
\par\medskip
However, to conclude that shared randomness is no good, would be
premature. Indeed, as we will indicate here, one of its uses may be
to turn the awkward average error performance into maximal error bounds.
This is something nontrivial --- in contrast to the case of single--sender
coding where the two concepts are essentially equivalent ---,
for it is known that the \emph{maximal error} concept can
yield strictly smaller capacity regions than the
\emph{average error} condition~\cite{dueck:max:vs:ave}.
\par
Let us consider for simplicity a two--sender multiple access channel, with
a code of rate $R_i$ for sender $i$ ($i=1,2$), and assume that there is common
randomness of rate $R_i$ between the sender $i$ and the receiver: let
the messages be represented by integers $m_i\in\{0,\ldots,M_i-1\}$,
$M_i=\lceil 2^{nR_i} \rceil$, and the common randomness as uniformly distributed
random variables $X_i\in\{0,\ldots,M_i-1\}$ ($i=1,2$).
\par
Sender $i$ then uses the given code to encode the message $m_i$ as
$n_i=m_i+X_i\mod M_i$ and sends the codeword corresponding to $n_i$
through the channel. The receiver first uses the given code to decode
an estimate $\widetilde{n_i}$ of $n_i$ and then computes
$\widetilde{m_i}=\widetilde{n_i}-X_i\mod M_i$ as estimate for $m_i$
($i=1,2$). Clearly, the average error probability of the given code equals
the individual message error probability (and hence the maximum error
probability) of this scheme.
\par
While this (simple) scheme requires quite a lot of common randomness,
standard derandomisation techniques (see e.g.~the communication
complexity textbook~\cite{kushilevitz:nisan})
show that $O(\log n)$ bits suffice on block length $n$, at the cost
of increasing the error probability by a constant factor.
This in turn implies that using randomised encodings one can
make the maximal error capacity region equal to Ahlswede's
average error capacity region.

\end{document}